\documentclass[onecolumn,prc,superscriptaddress,unsortedaddress,showpacs,preprintnumbers,amsmath,amssymb]{revtex4}
\usepackage{graphicx}
\usepackage{dcolumn}
\usepackage{bm}
\usepackage{CJK}
\usepackage{CJKulem}
\usepackage{txfonts}
\usepackage[dvipdfmx,bookmarks=true,colorlinks,%
citecolor=blue,linkcolor=blue,anchorcolor=blue,filecolor=blue,urlcolor=blue,%
           ]{hyperref}

\def\beq{\begin{equation}}
\def\eeq{\end{equation}}
\def\bea{\begin{eqnarray}}
\def\eea{\end{eqnarray}}

\def\fun#1#2{\lower3.6pt\vbox{\baselineskip0pt\lineskip.9pt
  \ialign{$\mathsurround=0pt#1\hfil##\hfil$\crcr#2\crcr\sim\crcr}}}

\usepackage[scaled=0.92]{helvet}
\usepackage[T1]{fontenc}
\usepackage{textcomp}

\begin{document}
\preprint{}

\title{Beyond Wigner's Isobaric Multiplet Mass Equation:
Effect of Charge-Symmetry-Breaking Interaction and Coulomb
Polarization}

\author{J. M. Dong}\email[ ]{dongjm07@impcas.ac.cn}\affiliation{Institute of Modern Physics, Chinese
Academy of Sciences, Lanzhou 730000, China}
\author{J. Z. Gu}
\affiliation{China Institute of Atomic Energy, P. O. Box 275(10),
Beijing 102413, China}
\author{Y. H. Zhang}
\affiliation{Institute of Modern Physics, Chinese Academy of
Sciences, Lanzhou 730000, China}
\author{W. Zuo}\email[ ]{zuowei@impcas.ac.cn}
\affiliation{Institute of Modern Physics, Chinese Academy of
Sciences, Lanzhou 730000, China} \affiliation{School of Physics,
University of Chinese Academy of Sciences, Beijing 100049, China}
\author{L. J. Wang}
\affiliation{Department of Physics and Astronomy, University of
North Carolina, Chapel Hill, North Carolina, 27516-3255, USA}
\author{Yu. A. Litvinov}
\affiliation{GSI Helmholtzzentrum f\"{u}r Schwerionenforschung,
Planckstra{\ss}e 1, 64291 Darmstadt, Germany}
\author{Y. Sun}\email[ ]{sunyang@sjtu.edu.cn }
\affiliation{Institute of Modern Physics, Chinese Academy of
Sciences, Lanzhou 730000, China} \affiliation{School of Physics and
Astronomy, Shanghai Jiao Tong University, Shanghai 200240, China}
\affiliation{Collaborative Innovation Center of IFSA, Shanghai Jiao
Tong University, Shanghai 200240, China}

\date{\today}

\begin{abstract}
The quadratic form of the isobaric multiplet mass equation (IMME),
which was originally suggested by Wigner and has been generally
regarded as valid, is seriously questioned by recent high-precision
nuclear mass measurements. The usual resolution to this problem is
to add empirically the cubic and quartic $T_z$-terms to characterize
the deviations from the IMME, but finding the origin of these terms
remains an unsolved difficulty. Based on a strategy beyond the
Wigner's first-order perturbation, we derive explicitly the cubic
and quartic $T_z$-terms. These terms are shown to be generated by
the effective charge-symmetry breaking and charge-independent
breaking interactions in nuclear medium combined with the Coulomb
polarization effect. Calculations for the $sd$- and lower
$fp$-shells explore a systematical emergence of the cubic
$T_z$-term, suggesting a general deviation from the original IMME.
Intriguingly, the magnitude of the deviation exhibits an
oscillation-like behavior with mass number, modulated by the shell
effect.

\end{abstract}

\maketitle

\section{Introduction}\noindent
Shortly after the discovery of neutron, Heisenberg introduced
isospin to describe different charge states of
nucleon~\cite{Tisospin}. In this concept, proton ($p$) and neutron
($n$) are treated as an isospin $T=1/2$ doublet distinguished by
different projections $T_z(p)=-1/2$ and $T_z(n)=+1/2$. As one of the
most important predictions in nuclear physics, the isobaric
multiplet mass equation (IMME) proposed later by
Wigner~\cite{IMME1,IMME2} suggests that the mass excesses
$\text{ME}(A,T,T_z)$ of the nuclei belonging to an isospin multiplet
of mass number $A$ and total isospin $T$ follow a simple quadratic
equation
\begin{equation}
\text{ME}(A,T,T_{z})=a+bT_{z}+cT_{z}^{2},\label{AA}
\end{equation}
where $T_z=(N-Z)/2$ is the isospin projection, and the parameters
$a$, $b$ and $c$ are constants for a given multiplet. The elegant
IMME, though derived by using the 1st-order perturbation
approximation, has been widely employed to predict the unknown
masses of unstable neutron-deficient nuclei.

Since its establishment, the IMME is believed to be generally
valid~\cite{Review1}. With recent advances in radioactive beam
facilities, a wealth of exotic masses with increasing precision
became available~\cite{AME2017}. Unexpectedly large discrepancies
between the measured masses and the ones given by the quadratic form
of the IMME were observed~\cite{Mass9,Mass53,Mass20}. This calls for
an addition of a cubic term $dT_z^3$ or even a quartic term $eT_z^4$
to Eq.~(\ref{AA})~\cite{Cou1,Cou2,Data2014}. The origin of these
higher-order terms, which clearly lies beyond the original IMME of
Eq. (\ref{AA}), requires explanation.

Various mechanisms have been proposed to explain the deviations
found in individual cases, including the isospin mixing, the
high-order Coulomb effect, and the charge-dependent nucleon-nucleon
interaction~\cite{Mix1,Mix2,Mix3,Mix4}. However, to date there is no
consensus as to the origin of the observed large $dT_z^3$ terms. In
current shell-model calculations, isospin-nonconversing (INC)
interactions are determined through fitting to available
experimental data \cite{Brown5,Zuker02,Kaneko13,Lam13,Kaneko14},
which are however insufficient to explain the experimental $dT_z^3$
terms.

In a recent work~\cite{Dong2017}, we have laid out a theoretical
framework which considers the contributions of the charge-symmetry
breaking (CSB) and charge-independence breaking (CIB) components in
nuclear medium to the effective nucleon-nucleon force. We have found
that such effective INC interactions are density-dependent, and thus
can no longer be expressed as irreducible tensors as was done by
Wigner \cite{IMME1,IMME2}. This leads us to propose a generalized
IMME (GIMME)~\cite{Dong2017} to study the Nolen-Schiffer anomaly,
which is expressed as~\cite{Dong2017}
\begin{eqnarray}
\text{ME}(A,T,T_{z}) &=&a+\left( b_{c}+\Delta _{\text{nH}}+2a_{\text{sym,1}%
}^{\text{(CSB)}}(A, T_{z})\right) T_{z} \nonumber\\
&&+\left( c_{c}+\frac{4}{A}a_{\text{sym,2}}^{\text{(CIB)}}(A,
T_{z})\right) T_{z}^{2},  \label{GIMME}
\end{eqnarray}
where $\Delta _{\text{nH}} =0.782$ MeV is the neutron-hydrogen mass
difference. In Eq. (\ref{GIMME}), the two anticipated
isospin-symmetry breaking sources, the Coulomb and the nuclear
interactions, are clearly separated. The $T_{z}$-independent $b_c$
and $c_c$ coefficients are produced solely by the Coulomb
interaction, whereas the 1st (2nd)-order symmetry energy
$a_{\text{sym,1}}^{\text{(CSB)}}$
($a_{\text{sym,2}}^{\text{(CIB)}}$) originates from the CSB (CIB)
interaction in nuclear medium~\cite{Dong2017}.

In this Rapid Communication we apply the GIMME~\cite{Dong2017} to
explore the physics beyond the 1st-order perturbation in the IMME.
We show that the appearance of the high-order $T_z$ terms is a more
general phenomenon, and that the degree of the deviation to Eq.
(\ref{AA}), measured by the coefficient of the $T_z^3$ term, is
totally governed by shell effects with a remarkable $A$-dependence.

\section{$T=3/2$ isobaric quartets}\noindent
Within the 1st-order perturbation calculation, $|\alpha T
T_z\rangle$ is assumed to be eigenstate of the charge-independent
Hamiltonian $H_0$, with $\alpha$ for all additional quantum numbers
to label this state. The energy produced by the CSB and CIB
interactions is
given by $\langle \alpha TT_{z}|H_{\text{CSB+CIB}}|\alpha TT_{z}\rangle =2a_{\text{%
sym,1}}^{\text{(CSB)}}(A,T_{z})T_{z}+\frac{4}{A}a_{\text{sym,2}}^{\text{(CIB)%
}}(A,T_{z})T_{z}^{2}$ in the absence of the Coulomb
force~\cite{Dong2017}. In this case, the
$a_{\text{sym,1}}^{\text{(CSB)}}(A,T_{z})$ (and also
$a_{\text{sym,2}}^{\text{(CIB)}}(A,T_{z})$) is identical for all
members of an isobaric multiplet, as seen in the appendix. Thus, the
GIMME is reduced to the quadratic form of the IMME. However, if we
go beyond the 1st-order perturbation to calculate $\langle \alpha
T_{z}|H_{\text{CSB+CIB}}|\alpha T_{z}\rangle$ with inclusion of the
Coulomb polarization effect,
$a_{\text{sym,1}}^{\text{(CSB)}}(A,T_{z})$ (and also
$a_{\text{sym,2}}^{\text{(CIB)}}(A,T_{z})$) is no longer a
constant for a given isobaric multiplet. One may formally expand them to be  $a_{\text{sym,1}}^{\text{(CSB)}%
}(A,T_{z})T_{z}=a_{1}+b_{1}T_{z}+c_{1}T_{z}^{2}+d_{1}T_{z}^{3}$ and
$a_{\text{sym,2}}^{\text{(CIB)}%
}(A,T_{z})T_{z}^{2}=a_{2}+b_{2}T_{z}+c_{2}T_{z}^{2}+d_{2}T_{z}^{3}$
for the $T=3/2$ quartets. Equation (\ref{GIMME}) can then be
rearranged as
\begin{equation}
\text{ME}(A,T,T_{z})=a+bT_{z}+cT_{z}^{2}+dT_{z}^{3}, \label{ME}
\end{equation}
where the $d$ coefficient is explicitly expressed as
\begin{eqnarray}
d &=&-\frac{8\pi }{9}\int_{0}^{\infty
}r^{2}S_{1}^{\text{(CSB)}}(\rho
)\left( \delta \rho _{3/2}-\delta \rho _{-3/2}\right) dr  \notag \\
&&-\frac{8\pi }{9}\int_{0}^{\infty }\frac{r^{2}}{\rho (r)}S_{2}^{\text{(CIB)}%
}(\rho )\left( \delta \rho _{3/2}^{2}-\delta \rho _{-3/2}^{2}\right)
dr. \label{d4}
\end{eqnarray}
In Eq. (\ref{d4}), $\delta \rho _{T_{z}}=\left[ \rho _{n}(r)-\rho
_{p}(r)\right] ^{\text{core}}_{T_z}$ is the neutron- and
proton-density difference in the core of the nucleus with $T_z$. For
$T=T_z$ ($-T_z$) nuclei, where there are $|N-Z|$ excess neutrons
(protons), we call the rest nucleons (with an equal number of
protons and neutrons) the `core' in the discussion. The terms
$S_{1}^{\text{(CSB)}}$ and $S_{2}^{\text{(CIB)}}$ in Eq. (\ref{d4})
are density-dependent symmetry energies characterizing the INC
interactions, defined as
\begin{eqnarray}
S_{1}^{\text{(CSB)}}(\rho ) &=&\frac{\partial E^{\text{(CSB)}}(\rho ,\beta )%
}{\partial \beta }|_{\beta =0}, \\
S_{2}^{\text{(CIB)}}(\rho ) &=&\frac{1}{2}\frac{\partial ^{2}E^{\text{(CIB)}%
}(\rho ,\beta )}{\partial \beta ^{2}}|_{\beta =0}.
\end{eqnarray}
The above results are achieved based on the microscopic
Brueckner-Hartree-Fock approach with the AV18 (together with AV14)
interaction~\cite{Dong2017}, in which $E^{\text{(CSB)}}$
($E^{\text{(CIB)}}$) is the energy associated with the CSB (CIB)
interaction. The CIB interaction contributes about one order of
magnitude smaller than the CSB interaction. From Eq. (\ref{d4}), the
nonzero $d$ originates primarily from the 1st-order symmetry energy
difference between the $T_z=3/2$ core and $T_z=-3/2$ core. The
nucleonic density distributions inside the nucleus are calculated
within a Skyrme energy-density functional approach. Detailed
derivation of Eqs. (\ref{ME}, \ref{d4}) are given in the appendix.
We note that the $dT_{z}^{3}$ term is usually added empirically to
account for deviations from the quadratic form of the IMME, but
here, it is derived microscopically. The occurrence of the nonzero
$d$ coefficient is due to a combined effect of the CSB interaction
in nuclear medium together with the treatment of the
beyond-1st-order perturbation calculation (i.e., by including the
core polarization induced by the Coulomb force). Both of them are
indispensable. We thus conclude that although the high-order Coulomb
contribution to the $d$ value is generally believed to be small
\cite{COUH}, it is necessarily to be included because without it, as
we discussed in detail before Eq. (\ref{ME}), $d$ term would not
appear.

\begin{figure}[htbp]
\begin{center}
\includegraphics[width=0.48\textwidth]{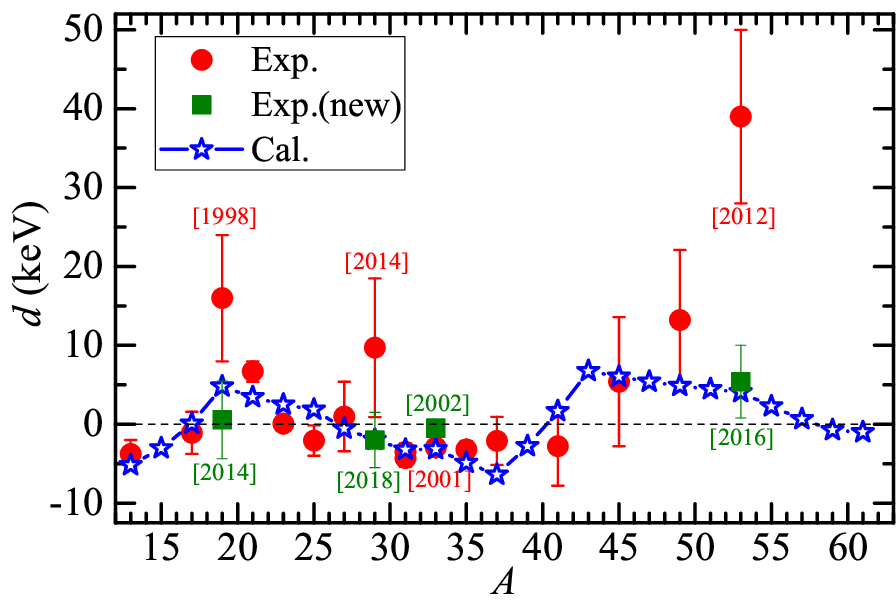}
\caption{(Color online) $d$ values for $T=3/2$ quartets calculated
with Eq. (\ref{d4}). The presented experimental data are from
Refs.~\cite{Data1998,Data2014} for $A=19$, \cite{Mass20} for $21$,
\cite{Data2014,Zhang2018} for $29$, \cite{Mass31a,Mass31b} for $31$,
\cite{Mass331,Mass332} for $33$, \cite{Mass35} for $35$,
\cite{Mass37} for $37$, \cite{Mass53} for $41-49$,
\cite{Mass53,Mass53-new} for $53$, and the remaining data are taken
from Ref.~\cite{Data2014}. The numbers in brackets refer to the
years of publication.}\label{fig1}
\end{center}
\end{figure}

In Fig.~\ref{fig1}, we show the $d$ values calculated with
Eq.~(\ref{d4}) for $T=3/2$ isobaric quartets. The SLy4
interaction~\cite{SLY} that satisfies specific constraints defined
in our previous work~\cite{Dong15} is employed to compute the
nucleonic density distributions. Remarkably, the results for the $d$
coefficients exhibit a clear $A$-dependence. It is striking that
across the $sd$ and lower $fp$ shells, the $d$ values show an
oscillation-like behavior with a minimum $\sim -6$ keV and maximum
$\sim 7$ keV. These occur across the magic numbers 8, 20, and 28,
implying the shell effect behind.

Experimental data seem to support the above prediction. In
Fig.~\ref{fig1}, we also plot the experimentally-extracted $d$
values for the isobaric quartets from the measured mass excesses via
$d=[$ME$(T_{z}=3/2) -$
ME$(T_{z}=-3/2)-3$ME$(T_{z}=1/2)+3$ME$(T_{z}=-1/2)]/6$. These
experimental $d$-values follow the predicted pattern well.

As one can see from Fig.~\ref{fig1}, the experimental $d$-values for
$A=31$~\cite{Mass31a,Mass31b}, $33$ \cite{Mass331,Mass332}, and
$35$~\cite{Mass35} multiplets have small uncertainties. Our
calculations reproduce these data. With the new mass of $^{29}$S
measured with the isochronous mass spectrometry technique in CSRe
recently \cite{Zhang2018}, the IMME is shown to be revalidated for
$A=29$, which is supported by our calculation. However, the validity
of the IMME represents only special cases of our general conclusion.

For the $T_z=3/2$ isobaric quartets, the first test in the
$fp$-shell with $A=45$, 49, and 53 indicated systematical deviations
from the IMME~\cite{Mass53}. Calculations based on two INC
Hamiltonians, the $f_{7/2}$ model space~\cite{Brown1} and the full
$pf$ model space with the GPFX1A interaction ~\cite{Brown2a,Brown2b}
plus the Ormand-Brown INC Hamiltonian~\cite{Brown5}, could not
reproduce the experimental data ~\cite{Mass53}. Our calculated $d$
values for $A=41$, 45, and 49 shown in Fig.~\ref{fig1} agree
qualitatively with experiment. For $A=53$, Ref.~\cite{Mass53}
initially reported a very large $d=39(11)$ keV. With a later
remeasurement of the IAS of $^{53}$Co~\cite{Mass53-new}, it was
reduced and agrees now well with our prediction (see
Fig.~\ref{fig1}). From our calculations, significant $d$-values are
expected in the $pf$-shell. Precision mass measurements of the
relevant nuclei are required. Especially interesting would be the
confirmation of the maximum $d$-value at $A=43$, for which the
identification and mass determination of the $T=3/2$ IAS in
$^{43}$Ti are needed.

The $A=21$ multiplet has been taken in Ref. \cite{Mass20} as an
example to show violations of the IMME in the $sd$-shell nuclei. The
universal $sd$ USDA and USDB isospin-conserving Hamiltonians
supplemented with an INC part yield too small $d$ values ($d=-0.3$
keV (USDA) and 0.3 keV (USDB) for the $A=21$, $J^\pi=5/2^+$
quartet~\cite{Mass20}). The valence-space calculations based on the
low-momentum two-nucleon and three-nucleon forces derived from the
chiral effective field theory~\cite{chiral} give $d=-38$ keV for the
$A=21$ quartet~\cite{Mass20}, which disagrees with the experimental
value 6.7(13) keV. Our result $d=3.4$ keV is close to the
experimental data.

\begin{figure}[htbp]
\begin{center}
\includegraphics[width=0.45\textwidth]{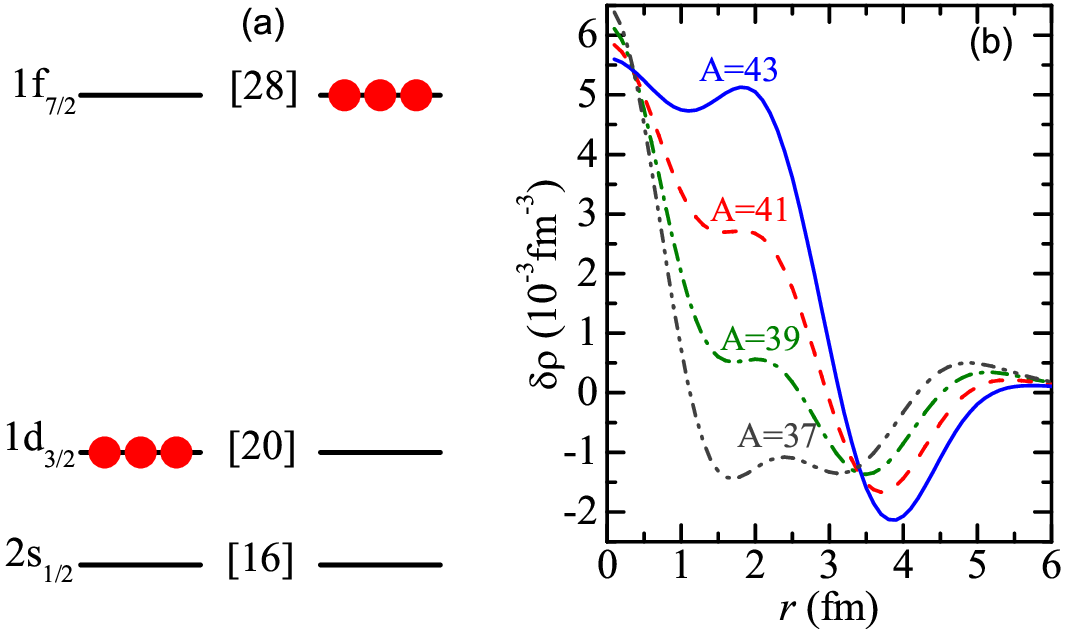}
\caption{(Color online) (a) Schematic illustration highlighting the
filling pattern of excess neutrons for the $T_z=T$ nucleus. (b) The
core density difference $\delta \rho _{3/2}-\delta \rho _{-3/2}$ for
$T=3/2$ isobaric quartets. $A=37$ and $A=43$ correspond to the left
and right panels of (a), respectively. }\label{fig2}
\end{center}
\end{figure}

We stress that the occurrence of nonzero $d$'s, which marks
deviations from the original IMME, is fundamental. The variation of
the $d$ coefficient is driven by the shell effect. We find that once
the excess neutrons in the $T_z=3/2$ of a multiplet fill a level
below (above) a large shell gap, as schematically illustrated in the
left (right) panel of Fig.~\ref{fig2}(a), the smallest (largest) $d$
value appears (see Fig. \ref{fig1}). For instance, the $T_z=T$
member of the $A=37$ ($A=43$) multiplet, $^{37}$Cl ($^{43}$Ca), has
three excess neutrons filling below (above) the $N=20$ shell gap.
When $A$ changes from 37 to 43, the excess neutrons (protons) for
$T_z=3/2$ ($T_z=-3/2$) member gradually occupy the upper $1f_{7/2}$
orbit. The neutrons (protons) in the core tend to be more loosely
(tightly) bound if they and the excess neutrons (protons) fill the
same (different) orbit(s), leading to the $A$-dependent differences
in neutron (proton) density of the core. As seen in
Fig.~\ref{fig2}(b), when $A$
changes from 37 to 43, $\delta \rho _{3/2}-\delta \rho _{-3/2}=\left[ \rho _{n}^{\text{core}%
}(r)-\rho _{p}(r)\right] _{T_{z}=3/2}+\left[ \rho
_{p}^{\text{core}}(r)-\rho _{n}(r)\right] _{T_{z}=-3/2}$ increases
significantly, particularly in the region of $r=1-3$ fm. This shell
effect is brought into the $d$ coefficient via the integral in
Eq.~(\ref{d4}). In the $T=3/2$ multiplets, the excess neutrons
(protons) in the $T_z=3/2$ ($T_z=-3/2$) nuclei occupy a level above
the $1p_{1/2}-1d_{5/2}$ or $1d_{3/2}-1f_{7/2}$ shell gap for the
$A=21$ or $A=45-53$ quartets, respectively, leading to a relatively
large violation of the IMME. This mechanism holds also true for the
strong breakdown of the $A = 9$ quartet since the excess neutrons
(protons) in the $T_z=3/2$ ($T_z=-3/2$) nuclei occupy a level above
the $1s_{1/2}-1p_{3/2}$ shell gap. However, the $A=9$ quartet is
excluded from Fig. \ref{fig1} for discussion because generally, the
Skyrme functional does not quantitatively apply to such light
nuclei. We conclude that the magnitude of the $d$ coefficient, which
measures the degree of deviation from the original IMME, depends on
the shell filling.

As shown in Fig.~\ref{fig1}, we predict a local maximum of $d$ for
the $A=19$ quartet. The existing experimental measurements are
divergent for the excitation energy of the IAS in
$^{19}$Ne~\cite{Data1998,Data2014}, leading to completely different
conclusions regarding the IMME. An experimental confirmation of this
energy is necessary.

\begin{figure}[htbp]
\begin{center}
\includegraphics[width=0.48\textwidth]{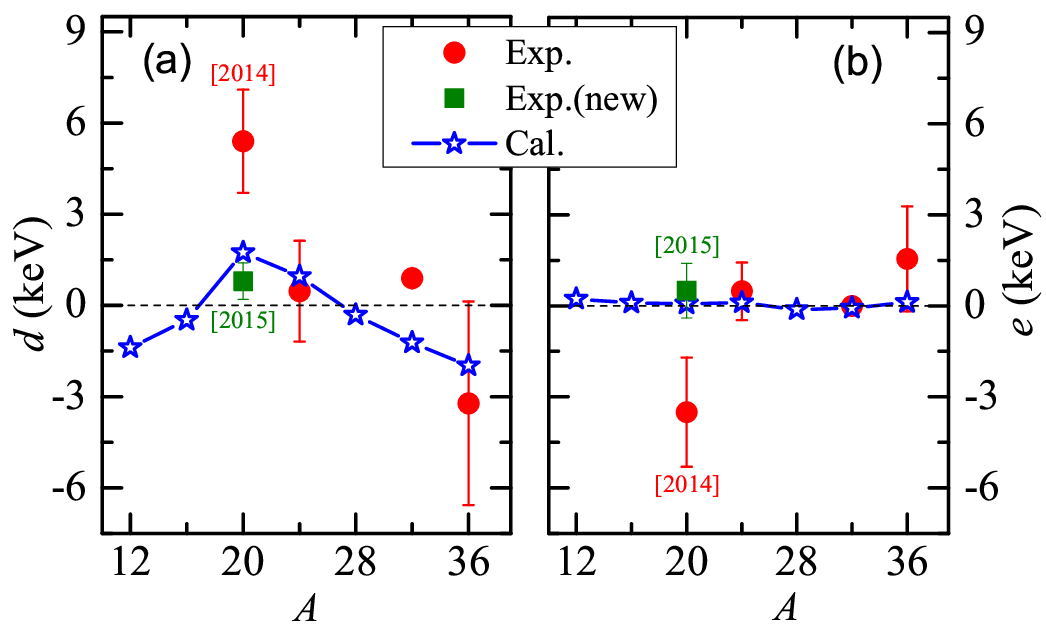}
\caption{(Color online) $d$ and $e$ values for $T=2$ quintets
calculated with Eqs. (\ref{d5}, \ref{e5}). The presented
experimental data are taken from Refs.~\cite{Mass20,Glassman} for
$A=20$, and the remaining data are from Ref.~\cite{Data2014}. The
numbers in brackets refer to the years of publication.}\label{fig3}
\end{center}
\end{figure}

\section{$T=2$ isobaric quintets}\noindent
With the similar derivation for $T=3/2$ quartets, the GIMME of Eq.
(\ref{GIMME}) is rewritten for $T=2$ quintets as
\begin{equation}
\text{ME}(A,T,T_{z})=a+bT_{z}+cT_{z}^{2}+dT_{z}^{3}+eT_{z}^{4},
\label{ME5}
\end{equation}
with the $d$ and $e$ coefficients given as
\begin{eqnarray}
d &=&-\frac{\pi }{4}\int_{0}^{\infty }r^{2}S_{1}^{\text{(CSB)}}(\rho
)\left(
\delta \rho _{2}-\delta \rho _{-2}\right) dr  \notag \\
&&-\frac{\pi }{4}\int_{0}^{\infty }\frac{r^{2}}{\rho (r)}S_{2}^{\text{(CIB)}%
}(\rho )\left( \delta \rho _{2}^{2}-\delta \rho _{-2}^{2}\right) dr,
\label{d5}
\\
e &=&-\frac{\pi }{64}\int_{0}^{\infty }dr\frac{r^{2}}{\rho (r)}S_{2}^{\text{%
(CIB)}}(\rho )\left( \delta \rho _{2}-\delta \rho _{-2}\right)
\times
\notag \\
&&\left[ 11\left( \delta \rho _{2}-\delta \rho _{-2}\right) +8\left(
\rho
_{n}^{\text{exc}}|_{T_{z}=2}+\rho _{p}^{\text{exc}}|_{T_{z}=-2}\right) %
\right],\label{e5}
\end{eqnarray}
where $\rho _{n}^{\text{exc}}|_{T_{z}=2}$ ($\rho
_{p}^{\text{exc}}|_{T_{z}=-2}$) is the density of the $|N-Z|$ excess
neutrons (protons) in the $T_z=T$ ($T_z=-T$) nucleus. The $e$
coefficient just originates from the CIB interaction, and the
detailed derivation is presented in the appendix.

In Fig.~\ref{fig3}, we compare the calculated $d$ and $e$ values for
$T=2$ quintets with the available experimental data. The
calculations suggest that the $d$ values are overall small, and the
magnitude does not exceed 2 keV. The calculated $d$ values for
$A=12-36$ in Fig. \ref{fig3}(a) show a similar pattern as the one
for $T=3/2$ quartets with a maximum at $A=20$. The occurrence of
this variation shares the same physical origin as that in the
$T=3/2$ quartets, namely the excess neutrons in the $T_z=T=2$ member
occupy a level above the $1p_{1/2}-1d_{5/2}$ shell gap, resulting in
a relatively larger $d$ value as compared with those of its
neighbors.

In 2014, a significant violation of the IMME for the $A=20$ quintet
was reported ~\cite{Mass20}. However, with a later measurement of
the excitation energy of the lowest $T=2$ state in $^{20}$Na, the
IMME was revalidated~\cite{Glassman}. Our calculated value of
$d=1.7$ keV is in agreement with the later experimental measurement
(see Fig.~\ref{fig3}(a)). The calculation presented in
Ref.~\cite{Mass20} deviates substantially from both measurements.

The data point for $A=32$ shows a deviation to our calculated value.
The studies of this quintet were carried out by several
collaborations~\cite{Mass320,Mass321,Mass322,Mass323,Mass324}, and
some of the measured data are controversial~\cite{Mass324}. We note
that both, the experimental and our predicted values, are small. The
calculated and experimental $e$ values are compatible with zero, and
they are in agreement to each other.

\section{Summary}\noindent
Based on our recently-proposed GIMME~\cite{Dong2017}, we have
established an isobaric multiplet mass equation that includes
high-order terms within a new strategy beyond the 1st-order
perturbation approximation. The explicit expression of $d$ (and also
$e$) coefficient which quantifies the deviation of the quadratic
form of the original IMME has been derived. The emergence of the
nonzero cubic $T_z$-term, and hence the violation of the quadratic
IMME, have basic roots. We have found that the
charge-symmetry-breaking interaction in nuclear medium,
characterized in our theory by the 1st-order symmetry energy,
combined with the core polarization effect primarily induced by
Coulomb force, are responsible for the breakdown of the quadratic
IMME. The effective charge-symmetry-breaking and
charge-independence-breaking interactions were extracted by
employing an {\it ab} initio method, i.e., the Brueckner theory with
bare AV18 and AV14 interactions without any phenomenological
adjustments. Remarkably, we found that the calculated $d$ values for
quartets and quintets follow an oscillation-like behavior throughout
the $sd$- and lower $fp$-shell regions as a consequence of the shell
effect, and the experimental $d$ values extracted from the measured
masses agree the oscillation pattern qualitatively. If all excess
neutrons in the $T_z=T$ nucleus of a multiplet fill a level above a
large shell gap, the breakdown of the quadratic IMME tends to be
strong. Therefore, it is straightforward to predict the nuclei where
the strong deviations from the original IMME are expected, which is
essential for guiding future experimental efforts.

\section*{Acknowledgements}\noindent
This work was supported by the National Natural Science Foundation
of China under Grants No. 11775276, No. 11435014, No. 11405223, No.
11675265, and No. 11575112, by the 973 Program of China under Grants
No. 2013CB834401 and No. 2013CB834405, by the National Key Program
for S\&T Research and Development (No. 2016YFA0400501,
2016YFA0400502), by the Youth Innovation Promotion Association of
Chinese Academy of Sciences, by the Helmholtz-CAS Joint Research
Group HCJRG-108, and by the European Research Council (ERC) under
the European Union's Horizon 2020 research and innovation program
(Grant agreement No. 682841 \textquotedblleft
ASTRUm\textquotedblright). Y.H.Z. acknowledges support by the
ExtreMe Matter Institute EMMI at the GSI Helmholtzzentrum f¨¹r
Schwerionenforschung, Darmstadt, Germany.

\section*{APPENDIX}\noindent
The 1st-order symmetry energy $a_{\text{sym,1}}^{\text{(CSB)}}(A,
T_{z})$ and the 2nd-order one $a_{\text{sym,2}}^{\text{(CIB)}}(A,
T_{z})$ for finite nuclei in Eq. (\ref{GIMME}) originate from the
CSB and CIB interactions in nuclear medium, respectively, which can
be expressed as~\cite{Dong2017}
\begin{eqnarray}
a_{\text{sym,1}}^{\text{(CSB)}}(A, T_{z})
&=&\frac{1}{IA}\int_{0}^{\infty
}4\pi r^{2}\rho (r)S_{1}^{\text{(CSB)}}(\rho )\beta (r)dr,  \label{GG} \\
a_{\text{sym,2}}^{\text{(CIB)}}(A, T_{z})
&=&\frac{1}{I^{2}A}\int_{0}^{\infty }4\pi r^{2}\rho
(r)S_{2}^{\text{(CIB)}}(\rho )\beta ^2 (r)dr,
\end{eqnarray}
where $I=(N-Z)/A=2T_z/A$ is the isospin asymmetry of a given
nucleus. $\beta(r) =(\rho _{n}(r)-\rho _{p}(r))/\rho(r)$ is the
local isospin asymmetry in which $\rho _{p}(r)$ and $\rho_{n}(r)$
are the proton and neutron density distributions inside the nucleus.
Here the symmetry energy coefficient is called simply as the
symmetry energy. The spherical-nuclei approximation is employed to
achieve a concise result. To achieve the
$a_{\text{sym,1}}^{\text{(CSB)}}(A, T_{z})$ and
$a_{\text{sym,2}}^{\text{(CIB)}}(A, T_{z})$, we should gain the
nucleonic density distributions $\rho_n$ and $\rho_p$ of an isobaric
analog state (IAS) whose $T$ is larger than $|T_z|$.

We assume $|\alpha T T_z\rangle$ is the eigenstate of the
charge-independent Hamiltonian $H_0$, with $\alpha$ for all
additional quantum numbers to label this state. In the first-order
perturbation, the energy produced by the CSB and CIB interactions is
given by $\langle \alpha TT_{z}|H_{\text{CSB+CIB}}|\alpha TT_{z}\rangle =2a_{\text{%
sym,1}}^{\text{(CSB)}}(A,T_{z})T_{z}+\frac{4}{A}a_{\text{sym,2}}^{\text{(CIB)%
}}(A,T_{z})T_{z}^{2}$ in the absence of Coulomb
force~\cite{Dong2017}. In this case, the wave function of the IAS
(with $T_z=T-1$) with $N-1$ neutrons and $Z+1$ protons ($N>Z$) is
obtained with
$|\text{IAS}\rangle=|T,T_{z}=T-1\rangle=\frac{1}{\sqrt{2T}}T_{-}|\text{0}\rangle$~\cite{IAS0}
rigidly, where $T_{-}=\underset{}{\sum }t_{-}(j)$ is the isospin
lowering operator, j $\in $ excess neutron orbits in
$|\text{0}\rangle$. $|\text{0}\rangle$ is the ground state of the
parent nucleus with $N$ neutrons and $Z$ protons belonging to a
multiplet with $T=T_z$. Thus, $(\rho _{n}+\rho
_{p})_{\text{IAS}}=(\rho _{n}+\rho _{p})_{\text{parent}}$ and $(\rho
_{n}-\rho _{p})_{\text{IAS}}=(1-\frac{1}{T})(\rho _{n}-\rho
_{p})_{\text{parent}}$ are obtained. The same situation also applies
to the IAS with $T_z=-(T-1)$. Accordingly, the
$a_{\text{sym,1}}^{\text{(CSB)}}(A,T_{z})$ (and also
$a_{\text{sym,2}}^{\text{(CIB)}}(A,T_{z})$) is identical for all
members of an isobaric multiplet. Therefore, they can be merged into
the $b_c$ and $c_c$, respectively. Namely, the GIMME is reduced to
the quadratic form of the IMME, and hence the IMME is not breakdown.

Here we treat this energy produced by CSB and CIB interactions
beyond the first-order perturbation approximation, i.e., calculate
the $\langle \alpha T_{z}|H_{\text{CSB+CIB}}|\alpha T_{z}\rangle$
instead of $\langle \alpha T T_{z}|H_{\text{CSB+CIB}}|\alpha T
T_{z}\rangle$, with the inclusion of Coulomb interaction.
Accordingly, the $a_{\text{sym,1}}^{\text{(CSB)}}(A,T_{z})$ (and
also $a_{\text{sym,2}}^{\text{(CIB)}}(A,T_{z})$) is no longer
a constant for a given isobaric multiplet. Yet, we can expand them as $a_{\text{sym,1}}^{\text{(CSB)}%
}(A,T_{z})T_{z}=a_{1}+b_{1}T_{z}+c_{1}T_{z}^{2}+d_{1}T_{z}^{3}$ and
$a_{\text{sym,2}}^{\text{(CIB)}%
}(A,T_{z})T_{z}^{2}=a_{2}+b_{2}T_{z}+c_{2}T_{z}^{2}+d_{2}T_{z}^{3}$
for $T=3/2$ isobaric quartets, and thus Eq. (\ref{GIMME}) is written
as
\begin{eqnarray}
\text{ME}(A,T,T_{z}) &=&a+\left( b_{c}+\Delta _{\text{nH}}\right)
T_{z}+2\left( a_{1}+b_{1}T_{z}+c_{1}T_{z}^{2}+d_{1}T_{z}^{3}\right)
\notag
\\
&&+c_{c}T_{z}^{2}+\frac{4}{A}\left(
a_{2}+b_{2}T_{z}+c_{2}T_{z}^{2}+d_{2}T_{z}^{3}\right) ,\notag \\
&=&\left( a+2a_{1}+\frac{4a_{2}}{A}\right) +\left( b_{c}+\Delta _{\text{nH}%
}+2b_{1}+\frac{4b_{2}}{A}\right) T_{z} \notag \\
&&+\left( c_{c}+2c_{1}+\frac{4c_{2}}{A}\right) T_{z}^{2}+\left( 2d_{1}+\frac{%
4d_{2}}{A}\right) T_{z}^{3}.
\end{eqnarray}
Therefore, if we introduce a new $a$ coefficient to replace the
$a+2a_{1}+\frac{4a_{2}}{A}$, and define the $b, c, d$ coefficients
by
\begin{eqnarray}
b &=&b_{c}+\Delta _{\text{nH}}+2b_{1}+\frac{4b_{2}}{A}, \nonumber\\
c &=&c_{c}+2c_{1}+\frac{4c_{2}}{A}, \nonumber \\
d &=&2d_{1}+\frac{4d_{2}}{A}
\end{eqnarray}
then the GIMME for quartets is rewritten as
\begin{equation}
\text{ME}(A,T,T_{z})=a+bT_{z}+cT_{z}^{2}+dT_{z}^{3}.
\end{equation}

\begin{figure}[htbp]
\begin{center}
\includegraphics[width=0.65\textwidth]{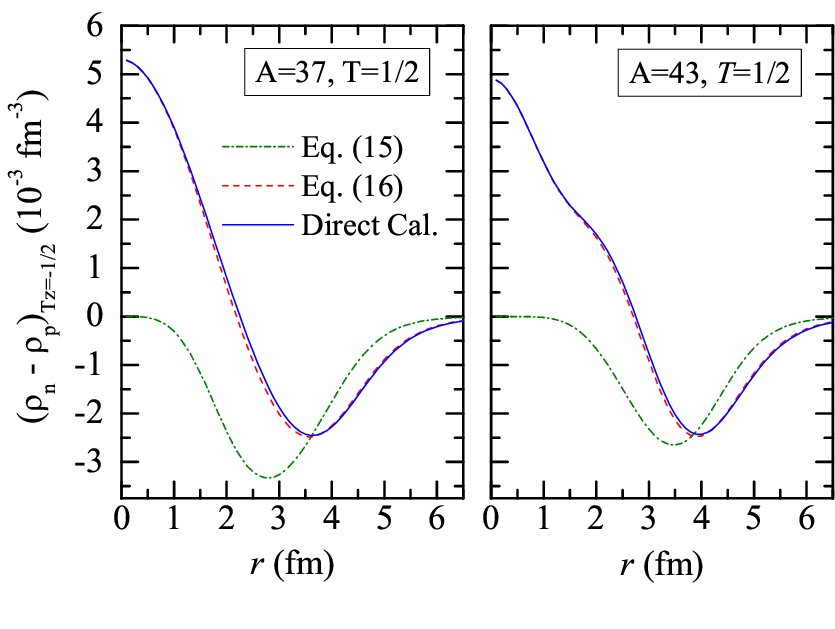}
\caption{The density difference between the neutron and proton,
i.e., $\rho _{n}-\rho _{p}$, for $A=37$ and $A=43$ isobaric doublets
as examples. The Eqs.~(\ref{old}, \ref{new}) are employed to
calculate the $\rho _{n}-\rho _{p}$ of $T_z=-1/2$ nuclei in the
framework of the Skyrme-Hartree-Fock method, and compared with the
direct calculation with the Skyrme-Hartree-Fock method.}
\label{test}
\end{center}
\end{figure}

The central task is to calculate the $d$ coefficient. It should be
stressed that, because of the core polarization induced by the
Coulomb force, the state of the $T_{-}|\text{0}\rangle/\sqrt{2T}$ is
not the exact description of the physical analog state ($T_z=T-1$).
The excess neutron density $\rho _{n,\text{parent}}^{\text{exc.}}$,
instead of $\rho _{n}-\rho _{p}$ should be used in the transition
density, as discussed in Ref.~\cite{IAS1}. Therefore, one obtains
\begin{equation}
(\rho _{n}-\rho _{p})_{_{T_{z}=T-1}}=\left( 1-\frac{1}{T}\right)
(\rho _{n}-\rho _{p})_{_{T_{z}=T}}+\frac{1}{T}\delta \rho
_{T_{z}=T},\label{old}
\end{equation}
where $\delta \rho (r) =\rho _{n}^{\text{core}}(r)-\rho _{p}(r)$ is
the neutron- and proton-density difference in the core (with an
equal number of protons and neutrons) in the $T_z=T$ nucleus. The
same situation also applies to the IAS with $T_z=-(T-1)$. In order
to improve the accuracy, we use the
\begin{equation}
(\rho _{n}-\rho _{p})_{_{T_{z}=T-1}}=\left( 1-\frac{1}{T}\right)
(\rho _{n}-\rho _{p})_{_{T_{z}=T}}+\frac{1}{2T}\left( \delta \rho
_{T_{z}=T}+\delta \rho _{T_{z}=T-1}\right).\label{new}
\end{equation}
In Fig. 1, we give two examples for $T=1/2$ doublet to show the
validity of such a treatment. Eq. (\ref{new}) works rather well, and
is much better than Eq. (\ref{old}). In a word, for the $T=3/2$
isobaric quartets, the density difference $\rho _{n}-\rho _{p}$ is
summarized as
\begin{equation}
\rho _{n}-\rho _{p}=%
\begin{cases}
\left( \rho _{n}-\rho _{p}\right) _{T_{z}=3/2}, & T_{z}=\frac{3}{2} \\
\frac{1}{3}\left( \rho _{n}-\rho _{p}\right)
_{T_{z}=3/2}+\frac{1}{3}\left(
\delta \rho _{3/2}+\delta \rho _{1/2}\right) , & T_{z}=\frac{1}{2} \\
\frac{1}{3}\left( \rho _{n}-\rho _{p}\right)
_{T_{z}=-3/2}+\frac{1}{3}\left(
\delta \rho _{-3/2}+\delta \rho _{-1/2}\right) , & T_{z}=-\frac{1}{2} \\
\left( \rho _{n}-\rho _{p}\right) _{T_{z}=-3/2}, & T_{z}=-\frac{3}{2}%
\end{cases}
\end{equation}
with $\delta \rho _{_{T_{z}}}=\left[ \rho _{n}^{\text{core}}(r)-\rho _{p}(r)%
\right] _{T_{z}}=\left[ \rho _{n}(r)-\rho _{p}(r)\right] _{T_{z}}^{\text{core%
}}$.

We assume that $\delta \rho _{3/2}, \delta \rho _{1/2}, \delta \rho
_{-1/2}, \delta \rho _{-3/2}$ are equidistant since the proton
number is equidistantly increasing from $T_z=3/2$ to $T_z=-3/2$
nuclei, and hence $\left( \delta \rho _{3/2}-\delta \rho
_{-3/2}\right) =3\left( \delta \rho _{1/2}-\delta \rho
_{-1/2}\right)$. Since we expand the first-order and the
second-order symmetry
energy produced by the CSB and CIB interactions as $a_{\text{sym,1}}^{\text{(CSB)}%
}(A,T_{z})T_{z}=a_{1}+b_{1}T_{z}+c_{1}T_{z}^{2}+d_{1}T_{z}^{3}$ and
$a_{\text{sym,2}}^{\text{(CIB)}%
}(A,T_{z})T_{z}^{2}=a_{2}+b_{2}T_{z}+c_{2}T_{z}^{2}+d_{2}T_{z}^{3}$,
the $d_1$ and $d_2$ are given as
\begin{eqnarray}
2d_{1} &=&\frac{1}{2}\left[ a_{\text{sym,1}}^{\text{(CSB)}}(A,\frac{3}{2}%
)+a_{\text{sym,1}}^{\text{(CSB)}}(A,-\frac{3}{2})-a_{\text{sym,1}}^{\text{%
(CSB)}}(A,\frac{1}{2})-a_{\text{sym,1}}^{\text{(CSB)}}(A,-\frac{1}{2})\right]
\notag \\
&=&\frac{1}{2}\frac{1}{3}\int_{0}^{\infty }4\pi r^{2}S_{1}^{\text{(CSB)}%
}(\rho )\left( \rho _{n}-\rho _{p}\right) _{T_{z}=3/2}dr  \notag \\
&&-\frac{1}{2}\frac{1}{3}\int_{0}^{\infty }4\pi r^{2}S_{1}^{\text{(CSB)}%
}(\rho )\left( \rho _{n}-\rho _{p}\right) _{T_{z}=-3/2}dr  \notag \\
&&-\frac{1}{2}\int_{0}^{\infty }4\pi r^{2}S_{1}^{\text{(CSB)}}(\rho
)\left[ \frac{1}{3}\left( \rho _{n}-\rho _{p}\right)
_{T_{z}=3/2}+\frac{1}{3}\left(
\delta \rho _{3/2}+\delta \rho _{1/2}\right) \right] dr  \notag \\
&&+\frac{1}{2}\int_{0}^{\infty }4\pi r^{2}S_{1}^{\text{(CSB)}}(\rho
)\left[ \frac{1}{3}\left( \rho _{n}-\rho _{p}\right)
_{T_{z}=-3/2}+\frac{1}{3}\left(
\delta \rho _{-3/2}+\delta \rho _{-1/2}\right) \right] dr  \notag \\
&=&\frac{1}{6}\int_{0}^{\infty }dr4\pi
r^{2}S_{1}^{\text{(CSB)}}(\rho )\cdot \bigg\{\left( \rho _{n}-\rho
_{p}\right) _{T_{z}=3/2}-\left( \rho
_{n}-\rho _{p}\right) _{T_{z}=-3/2}-  \notag \\
&&3\left[ \frac{1}{3}\left( \rho _{n}-\rho _{p}\right) _{T_{z}=3/2}+\frac{1}{%
3}\left( \delta \rho _{3/2}+\delta \rho _{1/2}\right) \right] +3\left[ \frac{%
1}{3}\left( \rho _{n}-\rho _{p}\right)
_{T_{z}=-3/2}+\frac{1}{3}\left(
\delta \rho _{-3/2}+\delta \rho _{-1/2}\right) \right] \bigg\}  \notag \\
&=&-\frac{1}{6}\int_{0}^{\infty }dr4\pi r^{2}S_{1}^{\text{(CSB)}}(\rho )%
\left[ \left( \delta \rho _{3/2}-\delta \rho _{-3/2}\right) +\left(
\delta
\rho _{1/2}-\delta \rho _{-1/2}\right) \right]  \notag\\
&=&-\frac{1}{6}\int_{0}^{\infty }dr4\pi r^{2}S_{1}^{\text{(CSB)}}(\rho )%
\left[ \left( \delta \rho _{3/2}-\delta \rho _{-3/2}\right) +\frac{1}{3}%
\left( \delta \rho _{3/2}-\delta \rho _{-3/2}\right) \right] \notag \\
&=&-\frac{8\pi }{9}\int_{0}^{\infty
}drr^{2}S_{1}^{\text{(CSB)}}(\rho )\left[ \left( \delta \rho
_{3/2}-\delta \rho _{-3/2}\right) \right] .
\end{eqnarray}%

\begin{eqnarray}
4d_{2} &=&\frac{1}{2}\left[ 3a_{\text{sym,2}}^{\text{(CIB)}}(A,\frac{3}{2}%
)-3a_{\text{sym,2}}^{\text{(CIB)}}(A,-\frac{3}{2})-a_{\text{sym,2}}^{\text{%
(CIB)}}(A,\frac{1}{2})+a_{\text{sym,2}}^{\text{(CIB)}}(A,-\frac{1}{2})\right]
\notag \\
&=&\frac{A}{8}\bigg\{\frac{3}{\left( \frac{3}{2}\right)
^{2}}\int_{0}^{\infty }4\pi r^{2}\frac{1}{\rho
(r)}S_{2}^{\text{(CIB)}}(\rho )\left( \rho
_{n}-\rho _{p}\right) _{T_{z}=3/2}^{2}dr  \notag \\
&&-\frac{3}{\left( -\frac{3}{2}\right) ^{2}}\int_{0}^{\infty }4\pi r^{2}%
\frac{1}{\rho (r)}S_{2}^{\text{(CIB)}}(\rho )\left( \rho _{n}-\rho
_{p}\right) _{T_{z}=-3/2}^{2}dr  \notag \\
&&-\frac{1}{\left( \frac{1}{2}\right) ^{2}}\int_{0}^{\infty }4\pi r^{2}\frac{%
1}{\rho (r)}S_{2}^{\text{(CIB)}}(\rho )\left[ \frac{1}{3}\left( \rho
_{n}-\rho _{p}\right) _{T_{z}=3/2}+\frac{1}{3}\left( \delta \rho
_{3/2}+\delta \rho _{1/2}\right) \right] ^{2}dr  \notag \\
&&+\frac{1}{\left( -\frac{1}{2}\right) ^{2}}\int_{0}^{\infty }4\pi r^{2}%
\frac{1}{\rho (r)}S_{2}^{\text{(CIB)}}(\rho )\left[
\frac{1}{3}\left( \rho _{n}-\rho _{p}\right)
_{T_{z}=-3/2}+\frac{1}{3}\left( \delta \rho
_{-3/2}+\delta \rho _{-1/2}\right) \right] ^{2}dr \bigg\}  \notag \\
&=&\frac{A}{8}\int_{0}^{\infty }dr4\pi r^{2}\frac{1}{\rho (r)}S_{2}^{\text{%
(CIB)}}(\rho )\cdot \bigg\{\frac{4}{3}\left( \rho _{n}-\rho
_{p}\right) _{T_{z}=3/2}^{2}-\frac{4}{3}\left( \rho _{n}-\rho
_{p}\right)
_{T_{z}=-3/2}^{2}  \notag \\
&&-4\left[ \frac{1}{3}\left( \rho _{n}-\rho _{p}\right) _{T_{z}=3/2}+\frac{1%
}{3}\left( \delta \rho _{3/2}+\delta \rho _{1/2}\right) \right]
^{2}+4\left[ \frac{1}{3}\left( \rho _{n}-\rho _{p}\right)
_{T_{z}=-3/2}+\frac{1}{3}\left(
\delta \rho _{-3/2}+\delta \rho _{-1/2}\right) \right] ^{2} \bigg\}  \notag \\
&=&\frac{A}{8}\int_{0}^{\infty }dr4\pi r^{2}\frac{1}{\rho (r)}S_{2}^{\text{%
(CIB)}}(\rho )\cdot \bigg\{\frac{8}{9}\left( \rho _{n}-\rho
_{p}\right) _{T_{z}=3/2}^{2}-\frac{8}{9}\left( \rho _{n}-\rho
_{p}\right)
_{T_{z}=-3/2}^{2}  \notag \\
&&-\frac{4}{9}\left[ \left( \delta \rho _{3/2}+\delta \rho
_{1/2}\right) ^{2}+2\left( \rho _{n}-\rho _{p}\right)
_{T_{z}=3/2}\left( \delta \rho
_{3/2}+\delta \rho _{1/2}\right) \right]   \notag \\
&&+\frac{4}{9}\left[ \left( \delta \rho _{-3/2}+\delta \rho
_{-1/2}\right) ^{2}+2\left( \rho _{n}-\rho _{p}\right)
_{T_{z}=-3/2}\left( \delta \rho
_{-3/2}+\delta \rho _{-1/2}\right) \right] \bigg\}  \notag \\
&=&\frac{A}{8}\int_{0}^{\infty }dr4\pi r^{2}\frac{1}{\rho (r)}S_{2}^{\text{%
(CIB)}}(\rho )\cdot \bigg\{\frac{8}{9}\left( \rho _{n}-\rho
_{p}\right) _{T_{z}=3/2}^{2}-\frac{8}{9}\left( \rho _{n}-\rho
_{p}\right) _{T_{z}=-3/2}^{2}-\frac{32}{27}\left( \delta \rho
_{3/2}^{2}-\delta \rho
_{-3/2}^{2}\right)   \notag \\
&&+\frac{8}{9}\left[ \left( \rho _{n}-\rho _{p}\right)
_{T_{z}=-3/2}\left( \frac{5}{3}\delta \rho _{-3/2}+\frac{1}{3}\delta
\rho _{3/2}\right) -\left(
\rho _{n}-\rho _{p}\right) _{T_{z}=3/2}\left( \frac{5}{3}\delta \rho _{3/2}+%
\frac{1}{3}\delta \rho _{-3/2}\right) \right] \bigg\}  \notag \\
&=&\frac{A}{8}\int_{0}^{\infty }dr4\pi r^{2}\frac{1}{\rho (r)}S_{2}^{\text{%
(CIB)}}(\rho )\cdot \bigg\{\frac{8}{9}\left[ \delta \rho
_{3/2}+\delta \rho
_{-3/2}\right] \cdot   \notag \\
&&\left[ \left( \rho _{n}-\rho _{p}\right) _{T_{z}=3/2}-\left( \rho
_{n}-\rho _{p}\right) _{T_{z}=-3/2}\right] -\frac{32}{27}\left(
\delta \rho
_{3/2}^{2}-\delta \rho _{-3/2}^{2}\right)   \notag \\
&&+\frac{8}{9}\left[ \left( \rho _{n}-\rho _{p}\right)
_{T_{z}=-3/2}\left( \frac{5}{3}\delta \rho _{-3/2}+\frac{1}{3}\delta
\rho _{3/2}\right) -\left(
\rho _{n}-\rho _{p}\right) _{T_{z}=3/2}\left( \frac{5}{3}\delta \rho _{3/2}+%
\frac{1}{3}\delta \rho _{-3/2}\right) \right] \bigg\}  \notag \\
&=&\frac{A}{8}\int_{0}^{\infty }dr4\pi r^{2}\frac{1}{\rho (r)}S_{2}^{\text{%
(CIB)}}(\rho )\left\{ -\frac{32}{27}\left( \delta \rho
_{3/2}^{2}-\delta \rho _{-3/2}^{2}\right) -\frac{16}{27}\left(
\delta \rho _{3/2}-\delta \rho _{-3/2}\right) \left( \delta \rho
_{3/2}+\delta \rho _{-3/2}\right) \right\}
\notag \\
&=&-\frac{2A}{9}\int_{0}^{\infty }dr4\pi r^{2}\frac{1}{\rho (r)}S_{2}^{\text{%
(CIB)}}(\rho )\left( \delta \rho _{3/2}^{2}-\delta \rho
_{-3/2}^{2}\right) .
\end{eqnarray}
Finally, the $d$ coefficient for the $T=3/2$ isobaric quartets takes
the form of
\begin{eqnarray}
d &=&2d_{1}+\frac{4d_{2}}{A}  \notag \\
&=&-\frac{8\pi }{9}\int_{0}^{\infty }r^{2}S_{1}^{\text{(CSB)}}(\rho
)\left(
\delta \rho _{3/2}-\delta \rho _{-3/2}\right) dr-\frac{8\pi }{9}%
\int_{0}^{\infty }r^{2}\frac{S_{2}^{\text{(CIB)}}(\rho )}{\rho
(r)}\left( \delta \rho _{3/2}^{2}-\delta \rho _{-3/2}^{2}\right) dr.
\end{eqnarray}

For $T=2$ isobaric quintets, similarly, the $\rho _{n}-\rho _{p}$ is
summarized as
\begin{equation}
\rho _{n}-\rho _{p}=%
\begin{cases}
\left( \rho _{n}-\rho _{p}\right) _{T_{z}=2}, & T_{z}=2 \nonumber \\
\frac{1}{2}\left( \rho _{n}-\rho _{p}\right)
_{T_{z}=2}+\frac{1}{4}\left(
\delta \rho _{2}+\delta \rho _{1}\right) , & T_{z}=1 \nonumber \\
\delta \rho _{0}, & T_{z}=0 \nonumber \\
\frac{1}{2}\left( \rho _{n}-\rho _{p}\right)
_{T_{z}=-2}+\frac{1}{4}\left(
\delta \rho _{-2}+\delta \rho _{-1}\right) , & T_{z}=-1 \nonumber \\
\left( \rho _{n}-\rho _{p}\right) _{T_{z}=-2}. & T_{z}=-2%
\end{cases} \label{density5}
\end{equation}
We expand the symmetry energy
$a_{\text{sym,1}}^{\text{(CSB)}}(A,T_{z})T_{z}$ and
$a_{\text{sym,2}}^{\text{(CIB)}}(A,T_{z})T_{z}^{2}$
as $a_{\text{sym,1}}^{\text{(CSB)}%
}(A,T_{z})T_{z}=a_{1}+b_{1}T_{z}+c_{1}T_{z}^{2}+d_{1}T_{z}^{3}+e_{1}T_{z}^{4}$
and
$a_{\text{sym,2}}^{\text{(CIB)}%
}(A,T_{z})T_{z}^{2}=a_{2}+b_{2}T_{z}+c_{2}T_{z}^{2}+d_{2}T_{z}^{3}+e_{2}T_{z}^{4}$,
and thus Eq. (\ref{GIMME}) is written as
\begin{eqnarray}
\text{ME}(A,T,T_{z}) &=&a+\left( b_{c}+\Delta _{\text{nH}}\right)
T_{z}+2\left(
a_{1}+b_{1}T_{z}+c_{1}T_{z}^{2}+d_{1}T_{z}^{3}+e_{1}T_{z}^{4}\right)
\notag
\\
&&+c_{c}T_{z}^{2}+\frac{4}{A}\left(
a_{2}+b_{2}T_{z}+c_{2}T_{z}^{2}+d_{2}T_{z}^{3}+e_{2}T_{z}^{4}\right)
,
\notag \\
&=&\left( a+2a_{1}+\frac{4a_{2}}{A}\right) +\left( b_{c}+\Delta _{\text{nH}%
}+2b_{1}+\frac{4b_{2}}{A}\right) T_{z}  \notag \\
&&+\left( c_{c}+2c_{1}+\frac{4c_{2}}{A}\right) T_{z}^{2}+\left( 2d_{1}+\frac{%
4d_{2}}{A}\right) T_{z}^{3}+\left( 2e_{1}+\frac{4e_{2}}{A}\right)
T_{z}^{4}.
\end{eqnarray}
The GIMME for quintets is reduced as
\begin{equation}
\text{ME}(A,T,T_{z})=a+bT_{z}+cT_{z}^{2}+dT_{z}^{3}+eT_{z}^{4}.
\label{ME5}
\end{equation}
with
\begin{eqnarray}
d &=&2d_{1}+\frac{4d_{2}}{A}, \\
e &=&2e_{1}+\frac{4e_{2}}{A}.\\ \nonumber
\end{eqnarray}

The $d_1, d_2, e_1, e_2$ coefficients are given by
\begin{eqnarray}
2d_{1} &=&\frac{1}{3}\left[ a_{\text{sym,1}}^{\text{(CSB)}}(A,2)+a_{\text{%
sym,1}}^{\text{(CSB)}}(A,-2)-a_{\text{sym,1}}^{\text{(CSB)}}(A,1)-a_{\text{%
sym,1}}^{\text{(CSB)}}(A,-1)\right]   \notag \\
&=&\frac{1}{12}\int_{0}^{\infty }4\pi r^{2}S_{1}^{\text{(CSB)}}(\rho
)\left( \rho _{n}-\rho _{p}\right)
_{T_{z}=2}dr-\frac{1}{12}\int_{0}^{\infty }4\pi
r^{2}S_{1}^{\text{(CSB)}}(\rho )\left( \rho _{n}-\rho _{p}\right)
_{T_{z}=-2}dr  \notag \\
&&-\frac{1}{6}\int_{0}^{\infty }4\pi r^{2}S_{1}^{\text{(CSB)}}(\rho
)\left[ \frac{1}{2}\left( \rho _{n}-\rho _{p}\right)
_{T_{z}=2}+\frac{1}{4}\left(
\delta \rho _{2}+\delta \rho _{1}\right) \right] dr  \notag \\
&&+\frac{1}{6}\int_{0}^{\infty }4\pi r^{2}S_{1}^{\text{(CSB)}}(\rho
)\left[ \frac{1}{2}\left( \rho _{n}-\rho _{p}\right)
_{T_{z}=-2}+\frac{1}{4}\left(
\delta \rho _{-2}+\delta \rho _{-1}\right) \right] dr \notag \\
&=&\frac{1}{12}\int_{0}^{\infty }dr4\pi
r^{2}S_{1}^{\text{(CSB)}}(\rho )\bigg\{\left( \rho _{n}-\rho
_{p}\right) _{T_{z}=2}-\left( \rho _{n}-\rho
_{p}\right) _{T_{z}=-2}-  \notag \\
&&2\left[ \frac{1}{2}\left( \rho _{n}-\rho _{p}\right) _{T_{z}=2}+\frac{1}{4}%
\left( \delta \rho _{2}+\delta \rho _{1}\right) \right] +2\left[ \frac{1}{2}%
\left( \rho _{n}-\rho _{p}\right) _{T_{z}=-2}+\frac{1}{4}\left(
\delta \rho
_{-2}+\delta \rho _{-1}\right) \right] \bigg\}  \notag \\
&=&-\frac{1}{24}\int_{0}^{\infty }dr4\pi r^{2}S_{1}^{\text{(CSB)}}(\rho )%
\left[ \delta \rho _{2}+\delta \rho _{1}-\delta \rho _{-2}-\delta \rho _{-1}%
\right]   \notag \\
&=&-\frac{1}{24}\int_{0}^{\infty }dr4\pi r^{2}S_{1}^{\text{(CSB)}}(\rho )%
\left[ \delta \rho _{2}-\delta \rho _{-2}+\frac{1}{2}\left( \delta
\rho
_{2}-\delta \rho _{-2}\right) \right]   \notag \\
&=&-\frac{\pi }{4}\int_{0}^{\infty }r^{2}S_{1}^{\text{(CSB)}}(\rho
)\left[ \delta \rho _{2}-\delta \rho _{-2}\right] dr.
\end{eqnarray}

\begin{eqnarray}
12e_{1} &=&a_{\text{sym,1}}^{\text{(CSB)}}(A,2)-a_{\text{sym,1}}^{\text{(CSB)%
}}(A,-2)-2a_{\text{sym,1}}^{\text{(CSB)}}(A,1)+2a_{\text{sym,1}}^{\text{(CSB)%
}}(A,-1)+3a_{\text{sym,1}}^{\text{(CSB)}}(A,T_{z})T_{z}|_{T_{z}=0}  \notag \\
&=&\frac{1}{4}\int_{0}^{\infty }4\pi r^{2}S_{1}^{\text{(CSB)}}(\rho
)\left( \rho _{n}-\rho _{p}\right)
_{T_{z}=2}dr+\frac{1}{4}\int_{0}^{\infty }4\pi
r^{2}S_{1}^{\text{(CSB)}}(\rho )\left( \rho _{n}-\rho _{p}\right)
_{T_{z}=-2}dr  \notag \\
&&-\int_{0}^{\infty }4\pi r^{2}S_{1}^{\text{(CSB)}}(\rho )\left[ \frac{1}{2}%
\left( \rho _{n}-\rho _{p}\right) _{T_{z}=2}+\frac{1}{4}\left(
\delta \rho
_{2}+\delta \rho _{1}\right) \right] dr  \notag \\
&&-\int_{0}^{\infty }4\pi r^{2}S_{1}^{\text{(CSB)}}(\rho )\left[ \frac{1}{2}%
\left( \rho _{n}-\rho _{p}\right) _{T_{z}=-2}+\frac{1}{4}\left(
\delta \rho
_{-2}+\delta \rho _{-1}\right) \right] dr+3\frac{1}{2T_{z}}%
T_{z}|_{T_{z}=0}\int_{0}^{\infty }4\pi
r^{2}S_{1}^{\text{(CSB)}}(\rho
)\delta \rho _{0}dr  \notag \\
&=&\frac{1}{4}\int_{0}^{\infty }dr4\pi
r^{2}S_{1}^{\text{(CSB)}}(\rho )\bigg\{\left( \rho _{n}-\rho
_{p}\right) _{T_{z}=2}+\left( \rho _{n}-\rho
_{p}\right) _{T_{z}=-2}-  \notag \\
&&\left[ 2\left( \rho _{n}-\rho _{p}\right) _{T_{z}=2}+\left( \delta
\rho _{2}+\delta \rho _{1}\right) \right] -\left[ 2\left( \rho
_{n}-\rho
_{p}\right) _{T_{z}=-2}+\left( \delta \rho _{-2}+\delta \rho _{-1}\right) %
\right] +6\delta \rho _{0}\bigg\}  \notag \\
&=&-\pi \int_{0}^{\infty }drr^{2}S_{1}^{\text{(CSB)}}(\rho
)\bigg\{\left[ \left( \rho _{n}-\rho _{p}\right) _{T_{z}=2}+\left(
\delta \rho _{2}+\delta \rho _{1}\right) \right] +\left[ \left( \rho
_{n}-\rho _{p}\right) _{T_{z}=-2}+\left( \delta \rho _{-2}+\delta
\rho _{-1}\right) \right]
-6\delta \rho _{0}\bigg\}  \notag \\
&=&-\pi \int_{0}^{\infty }drr^{2}S_{1}^{\text{(CSB)}}(\rho
)\bigg\{\left[ \left( \rho _{n}-\rho _{p}\right) _{T_{z}=2}+\left(
\rho _{n}-\rho _{p}\right) _{T_{z}=-2}\right] +\left[ \left( \delta
\rho _{2}+\delta \rho _{1}\right) +\left( \delta \rho _{-2}+\delta
\rho _{-1}\right) \right] -6\delta \rho
_{0}\bigg\}  \notag \\
&=&-\pi \int_{0}^{\infty }drr^{2}S_{1}^{\text{(CSB)}}(\rho )\left[
\left( \delta \rho _{2}+\delta \rho _{-2}\right) +\left( \delta \rho
_{2}+\delta \rho _{1}\right) +\left( \delta \rho _{-2}+\delta \rho
_{-1}\right) -6\delta
\rho _{0}\right]   \notag \\
&=&0.
\end{eqnarray}
\begin{eqnarray}
6d_{2} &=&2a_{\text{sym,2}}^{\text{(CIB)}}(A,2)-2a_{\text{sym,2}}^{\text{%
(CIB)}}(A,-2)-a_{\text{sym,2}}^{\text{(CIB)}}(A,1)+a_{\text{sym,2}}^{\text{%
(CIB)}}(A,-1)  \notag \\
&=&\frac{A}{4}\bigg\{\frac{2}{4}\int_{0}^{\infty }4\pi r^{2}\frac{1}{\rho (r)}%
S_{2}^{\text{(CIB)}}(\rho )\left( \rho _{n}-\rho _{p}\right)
_{T_{z}=2}^{2}dr-\frac{2}{4}\int_{0}^{\infty }4\pi r^{2}\frac{1}{\rho (r)}%
S_{2}^{\text{(CIB)}}(\rho )\left( \rho _{n}-\rho _{p}\right)
_{T_{z}=-2}^{2}dr  \notag \\
&&-\int_{0}^{\infty }4\pi r^{2}\frac{1}{\rho (r)}S_{2}^{\text{(CIB)}}(\rho )%
\left[ \frac{1}{2}\left( \rho _{n}-\rho _{p}\right) _{T_{z}=2}+\frac{1}{4}%
\left( \delta \rho _{2}+\delta \rho _{1}\right) \right] ^{2}dr  \notag \\
&&+\int_{0}^{\infty }4\pi r^{2}\frac{1}{\rho (r)}S_{2}^{\text{(CIB)}}(\rho )%
\left[ \frac{1}{2}\left( \rho _{n}-\rho _{p}\right) _{T_{z}=-2}+\frac{1}{4}%
\left( \delta \rho _{-2}+\delta \rho _{-1}\right) \right] ^{2}dr \bigg\}  \notag \\
&=&\frac{A}{4}\int_{0}^{\infty }dr4\pi r^{2}\frac{1}{\rho (r)}S_{2}^{\text{%
(CIB)}}(\rho )\cdot \bigg\{\frac{1}{2}\left( \rho _{n}-\rho
_{p}\right) _{T_{z}=2}^{2}-\frac{1}{2}\left( \rho _{n}-\rho
_{p}\right) _{T_{z}=-2}^{2}
\notag \\
&&-\left[ \frac{1}{2}\left( \rho _{n}-\rho _{p}\right) _{T_{z}=2}+\frac{1}{4}%
\left( \delta \rho _{2}+\delta \rho _{1}\right) \right] ^{2}+\left[ \frac{1}{%
2}\left( \rho _{n}-\rho _{p}\right) _{T_{z}=-2}+\frac{1}{4}\left(
\delta
\rho _{-2}+\delta \rho _{-1}\right) \right] ^{2}\bigg\}  \notag \\
&=&\frac{A}{16}\int_{0}^{\infty }dr4\pi r^{2}\frac{1}{\rho (r)}S_{2}^{\text{%
(CIB)}}(\rho )\bigg\{\left[ \left( \rho _{n}-\rho _{p}\right)
_{T_{z}=2}^{2}-\left( \rho _{n}-\rho _{p}\right)
_{T_{z}=-2}^{2}\right] -
\notag \\
&&\frac{1}{4}\left[ \left( \delta \rho _{2}+\delta \rho _{1}\right)
^{2}-\left( \delta \rho _{-2}+\delta \rho _{-1}\right) ^{2}\right]
-\left( \rho _{n}-\rho _{p}\right) _{T_{z}=2}\left( \delta \rho
_{2}+\delta \rho _{1}\right) +\left( \rho _{n}-\rho _{p}\right)
_{T_{z}=-2}\left( \delta \rho
_{-2}+\delta \rho _{-1}\right) \bigg\}  \notag \\
&=&\frac{A}{16}\int_{0}^{\infty }dr4\pi r^{2}\frac{1}{\rho (r)}S_{2}^{\text{%
(CIB)}}(\rho )\left\{ -\frac{3}{4}\left( \delta \rho _{2}^{2}-\delta
\rho _{-2}^{2}\right) -\frac{3}{4}\left( \delta \rho _{2}+\delta
\rho _{-2}\right) \left( \delta \rho _{2}-\delta \rho _{-2}\right)
\right\}
\notag \\
&=&\frac{A}{16}\int_{0}^{\infty }dr4\pi r^{2}\frac{1}{\rho (r)}S_{2}^{\text{%
(CIB)}}(\rho )\left\{ -\frac{3}{2}\left( \delta \rho _{2}^{2}-\delta
\rho
_{-2}^{2}\right) \right\}   \notag \\
&=&-\frac{3A}{32}\int_{0}^{\infty }dr4\pi r^{2}\frac{1}{\rho (r)}S_{2}^{%
\text{(CIB)}}(\rho )\left( \delta \rho _{2}^{2}-\delta \rho
_{-2}^{2}\right) .
\end{eqnarray}

\begin{eqnarray}
6e_{2} &=&a_{\text{sym,2}}^{\text{(CIB)}}(A,2)+a_{\text{sym,2}}^{\text{(CIB)}%
}(A,-2)-a_{\text{sym,2}}^{\text{(CIB)}}(A,1)-a_{\text{sym,2}}^{\text{(CIB)}%
}(A,-1)+\frac{3}{2}a_{\text{sym,2}}^{\text{(CIB)}%
}(A,T_{z})T_{z}^{2}|_{T_{z}=0}  \notag \\
&=&\frac{A}{4}\bigg\{\frac{1}{4}\int_{0}^{\infty }4\pi r^{2}\frac{1}{\rho (r)}%
S_{2}^{\text{(CIB)}}(\rho )\left( \rho _{n}-\rho _{p}\right)
_{T_{z}=2}^{2}dr+\frac{1}{4}\int_{0}^{\infty }4\pi r^{2}\frac{1}{\rho (r)}%
S_{2}^{\text{(CIB)}}(\rho )\left( \rho _{n}-\rho _{p}\right)
_{T_{z}=-2}^{2}dr  \notag \\
&&-\int_{0}^{\infty }\frac{4\pi r^{2}}{\rho (r)}S_{2}^{\text{(CIB)}}(\rho )%
\left[ \frac{1}{2}\left( \rho _{n}-\rho _{p}\right) _{T_{z}=2}+\frac{1}{4}%
\left( \delta \rho _{2}+\delta \rho _{1}\right) \right] ^{2}dr-  \notag \\
&&\int_{0}^{\infty }\frac{4\pi r^{2}}{\rho (r)}S_{2}^{\text{(CIB)}}(\rho )%
\left[ \frac{1}{2}\left( \rho _{n}-\rho _{p}\right) _{T_{z}=-2}+\frac{1}{4}%
\left( \delta \rho _{-2}+\delta \rho _{-1}\right) \right] ^{2}dr+\frac{3}{%
2T_{z}^{2}}T_{z}^{2}|_{T_{z}=0}\int_{0}^{\infty }4\pi r^{2}\frac{1}{\rho (r)}%
S_{2}^{\text{(CIB)}}(\rho )\delta \rho _{0}^{2}dr \bigg\}  \notag \\
&=&\frac{A}{16}\int_{0}^{\infty }dr4\pi r^{2}\frac{1}{\rho (r)}S_{2}^{\text{%
(CIB)}}(\rho )\cdot \bigg\{\left( \rho _{n}-\rho _{p}\right)
_{T_{z}=2}^{2}+\left( \rho _{n}-\rho _{p}\right) _{T_{z}=-2}^{2}  \notag \\
&&-\left[ \left( \rho _{n}-\rho _{p}\right)
_{T_{z}=2}+\frac{1}{2}\left( \delta \rho _{2}+\delta \rho
_{1}\right) \right] ^{2}-\left[ \left( \rho _{n}-\rho _{p}\right)
_{T_{z}=-2}+\frac{1}{2}\left( \delta \rho _{-2}+\delta
\rho _{-1}\right) \right] ^{2}+6\delta \rho _{0}^{2}\bigg\}  \notag \\
&=&-\frac{A}{16}\int_{0}^{\infty }dr4\pi r^{2}\frac{1}{\rho (r)}S_{2}^{\text{%
(CIB)}}(\rho )\cdot \bigg\{\frac{1}{4}\left( \delta \rho _{2}+\delta
\rho _{1}\right) ^{2}+\left( \rho _{n}-\rho _{p}\right)
_{T_{z}=2}\left( \delta
\rho _{2}+\delta \rho _{1}\right)  \nonumber \\
&&+\frac{1}{4}\left( \delta \rho _{-2}+\delta \rho _{-1}\right)
^{2}+\left( \rho _{n}-\rho _{p}\right) _{T_{z}=-2}\left( \delta \rho
_{-2}+\delta \rho
_{-1}\right) -6\delta \rho _{0}^{2}\bigg\}  \notag \\
&=&-\frac{A}{64}\int_{0}^{\infty }dr4\pi r^{2}\frac{1}{\rho (r)}S_{2}^{\text{%
(CIB)}}(\rho )\cdot \bigg\{\left( 2\delta \rho _{2}-\frac{\delta
\rho _{2}-\delta \rho _{-2}}{4}\right) ^{2}+4\left( \rho _{n}-\rho
_{p}\right)
_{T_{z}=2}\left( 2\delta \rho _{2}-\frac{\delta \rho _{2}-\delta \rho _{-2}}{%
4}\right)   \notag \\
&&+\left( 2\delta \rho _{-2}+\frac{\delta \rho _{2}-\delta \rho _{-2}}{4}%
\right) ^{2}+4\left( \rho _{n}-\rho _{p}\right) _{T_{z}=-2}\left(
2\delta \rho _{-2}+\frac{\delta \rho _{2}-\delta \rho
_{-2}}{4}\right) -24\delta
\rho _{0}^{2}\bigg\}  \notag \\
&=&-\frac{A}{64}\int_{0}^{\infty }dr4\pi r^{2}\frac{1}{\rho (r)}S_{2}^{\text{%
(CIB)}}(\rho )\cdot \bigg\{\left( \frac{25}{8}\delta \rho _{2}^{2}+\frac{25}{8}%
\delta \rho _{-2}^{2}+\frac{7}{4}\delta \rho _{2}\delta \rho
_{-2}\right) +
\notag \\
&&4\left( \rho _{n}-\rho _{p}\right) _{T_{z}=2}\left[ \left( \delta
\rho _{2}+\delta \rho _{-2}\right) +\frac{3}{4}\left( \delta \rho
_{2}-\delta
\rho _{-2}\right) \right] +4\left( \rho _{n}-\rho _{p}\right) _{T_{z}=-2}%
\left[ \left( \delta \rho _{2}+\delta \rho _{-2}\right)
-\frac{3}{4}\left( \delta \rho _{2}-\delta \rho _{-2}\right) \right]
-24\delta \rho _{0}^{2}\bigg\}
\notag \\
&=&-\frac{3A}{64}\int_{0}^{\infty }dr4\pi r^{2}\frac{1}{\rho (r)}S_{2}^{%
\text{(CIB)}}(\rho )\left\{ \frac{3}{8}\left( \delta \rho
_{2}-\delta \rho _{-2}\right) ^{2}+\left( \delta \rho _{2}-\delta
\rho _{-2}\right) \left[ \left( \rho _{n}-\rho _{p}\right)
_{T_{z}=2}-\left( \rho _{n}-\rho _{p}\right) _{T_{z}=-2}\right]
\right\} .
\end{eqnarray}
Finally, we obtain the $d$, $e$ coefficients for the $T=2$ quintets
taking the form of
\begin{eqnarray}
d &=&2d_{1}+\frac{4d_{2}}{A}  \notag \\
&=&-\frac{\pi }{4}\int_{0}^{\infty }r^{2}S_{1}^{\text{(CSB)}}(\rho
)\left(
\delta \rho _{2}-\delta \rho _{-2}\right) dr-\frac{\pi }{4}\int_{0}^{\infty }%
\frac{r^{2}}{\rho (r)}S_{2}^{\text{(CIB)}}(\rho )\left( \delta \rho
_{2}^{2}-\delta \rho _{-2}^{2}\right) dr, \\
e &=&2e_{1}+\frac{4e_{2}}{A}  \notag \\
&=&-\frac{\pi }{64}\int_{0}^{\infty }dr\frac{r^{2}}{\rho (r)}S_{2}^{\text{%
(CIB)}}(\rho )\left( \delta \rho _{2}-\delta \rho _{-2}\right) \cdot
\left[
11\left( \delta \rho _{2}-\delta \rho _{-2}\right) +8\left( \rho _{n}^{\text{%
exc}}|_{T_{z}=2}+\rho _{p}^{\text{exc}}|_{T_{z}=-2}\right) \right] ,
\end{eqnarray}
where $\rho _{n}^{\text{exc}}|_{T_{z}=2}$ ($\rho
_{p}^{\text{exc}}|_{T_{z}=-2}$) is the density of the $|N-Z|$ excess
neutrons (protons) in the $T_z=T$ ($T_z=-T$) nucleus.

\end{document}